\documentclass[pra,twocolumn,letterpaper,amsfonts]{revtex4}

\usepackage{graphicx}

\usepackage{amsthm}

\usepackage{amsfonts}

\usepackage{amssymb}

\usepackage{bm}

\begin{document}

\bibliographystyle{apsrev}

\newcommand{\Bra}[1]{\ensuremath{\left \langle #1 \right |}}
\newcommand{\Ket}[1]{\ensuremath{\left | #1 \right \rangle}}
\newcommand{\BraKet}[2]{\ensuremath{\left \langle #1 \right | \left. #2 \right \rangle}}
\newcommand{\tr}{\ensuremath{\mbox{Tr}}}
\renewcommand{\H}{\ensuremath{\mathcal{H}}}
\newcommand{\var}{\ensuremath{ \mbox{var}}}
\newcommand{\hvar}[1]{\ensuremath{ \hat{\mbox{var}} \left ( #1 \right )}}

\title{Measuring Polynomial Invariants of Multi-Party Quantum States}
\author{M. S. Leifer}
\email[Email: ]{Matt.Leifer@bristol.ac.uk}
\author{N. Linden}
\affiliation{Dept. of Mathematics, University of Bristol, University Walk,
Bristol, BS8 1TW, UK}
\author{A. Winter}
\affiliation{Dept. of Computer Science, University of Bristol, Merchant
Venturers Building, Woodland Road, Bristol, BS8 1UB, UK}
\date{\today}

\begin{abstract}
We present networks for directly estimating the polynomial invariants of
multi-party quantum states under local transformations.  The structure of these networks is closely related to the
structure of the invariants themselves and this lends a physical interpretation
to these otherwise abstract mathematical quantities.  Specifically, our networks
estimate the invariants under local unitary (LU) transformations and under stochastic
local operations and classical communication (SLOCC).  Our networks can estimate
the LU invariants for multi-party states, where each party can have a Hilbert
space of arbitrary dimension and the SLOCC invariants for multi-qubit states.
We analyze the statistical efficiency of our networks compared to methods based
on estimating the state coefficients and calculating the invariants.
\end{abstract}

\maketitle

\section{Introduction}

\label{Intro}

Entanglement is a key resource in quantum information and computation since it
can be used to perform tasks such as teleportation, super-dense coding and key
distribution.  Therefore, it is important to find ways of classifying and
quantifying the entanglement properties of quantum states.  Central to this is
the idea that locally invariant quantities can be used to characterize
entanglement.  Invariants under Local Unitary (LU) and more general
transformations, such as Stochastic Local Operations and Classical Communication
(SLOCC), have been extensively studied in this context \cite{Multi1, Multi2,
Multi3, Sud1, Sud2, GrasslInv, VerMulti1, VerMulti2, VerMulti3, JaegMultiStoke,
JaegLorentz, JaegEnt}.

However, invariants are rather abstract mathematical objects and it is
natural to ask whether any physical meaning can be given to them.  One way of
doing this is to investigate how these quantities might be measured given a
number of copies of an unknown state.  This could be done by simply
measuring the coefficients of the state and then calculating the invariants.
However, finding procedures to measure the invariants directly may be more efficient and also lends the invariants a physical interpretation as ``collective observables'' of the state.

For bipartite pure states, the Schmidt coefficients are a complete set of LU
invariants and optimal protocols for measuring them were given in
\cite{AcinEntEst}.  Also, in \cite{HorEntEst} a method was given for estimating
the polynomial SLOCC invariants of a general two-qubit state.

In this paper we present networks for estimating two classes of polynomial
invariants for multi-party states:  the LU invariants for multi-party states
with arbitrary local Hilbert space dimension and the SLOCC invariants for
multi-qubit states.  In both cases, the protocol works for both pure and mixed
states.   In particular, the structure of the networks reflects the structure of
the invariants in a very simple way.

In \S\ref{PolyLU}, we review the construction of local invariants under LU
transformations and in \S\ref{MeasLU}, the networks for measuring these
invariants are presented.  We then turn to invariants under SLOCC
transformations, reviewing their construction in \S\ref{PolySLOCC} and
presenting networks to measure them in \S\ref{MeasSLOCC}.  In order to construct
the networks for SLOCC invariants we make use of the Structural Physical
Approximation (SPA) to non-physical maps introduced in \cite{HorSPA}.  The
relevant details of this are presented in \S\ref{SPA}.  Finally, in \S\ref{Eval}
we evaluate estimation protocols based on our networks by comparing them to
simple techniques based on estimating the state coefficients.

\section{Polynomial Invariants under LU transformations}

\label{PolyLU}

\subsection{Pure states}

Two $n$-party pure states $\Ket{\psi}, \Ket{\psi'} \in \bigotimes_{j=1}^n
\mathbb{C}^{d_j}$ are equivalent under LU transformations if
\begin{equation}
\label{Poly:LUAction}
\Ket{\psi'} = U_1 \otimes U_2 \otimes \ldots \otimes U_n \Ket{\psi}
\end{equation}
where $U_j \in U(d_j)$ is a unitary operation acting on the Hilbert space of the $j$th party.
States on the same orbit under this action have the same entanglement
properties.  Given a particular state, we might be interested in determining
which orbit it belongs to.  This can be done by establishing a canonical
point on each orbit, such as the Schmidt form for bipartite states.  However,
canonical forms rapidly become more complicated as the number of parties is increased.
Alternatively, we can construct polynomial functions of the state coefficients
that are invariant on each orbit.  Theorems from invariant theory guarantee that
a finite set of such polynomials is enough to distinguish the generic orbits
under this action.  We now review the construction of such a set.

\subsubsection{One party}

Consider the state $\Ket{\psi} =
\sum_{i=1}^d \alpha^i \Ket{i}$ in a single party Hilbert space $\bm{C}^d$, where
$\{ \Ket{i} \}$ is an orthonormal basis.  The only independent invariant under unitary
transformations of this state is the norm $\BraKet{\psi}{\psi}$.
This may be written as
\begin{equation}
\BraKet{\psi}{\psi} = \sum_i \alpha^i \alpha^*_i  = \sum_{i,j} \alpha^i
\delta_i^j \alpha^*_j
\end{equation}
where $\delta_i^j$ is the Kronecker delta.  $\delta_i^j$ is the $U(d)$
invariant tensor and invariants for larger numbers of parties are formed by
similar contractions of the state coefficients with their complex conjugates.

\subsubsection{Two qubits}

As an example, consider a two-qubit state $\Ket{\psi} = \sum_{i,j = 0}^1 \alpha^{ij}
\Ket{ij}$.  There is only one independent quadratic invariant, which is simply
the norm of the state.  However, at quartic order we find the following
invariant, which is functionally independent of the norm
\begin{equation}
\label{Poly:2qqLU}
\begin{array}{lll}
J & = & \sum \alpha^{i_1 j_1} \alpha^{i_2 j_2} \delta_{i_1}^{i_3} \delta_{i_2}^{i_4}
\delta_{j_1}^{j_4} \delta_{j_2}^{j_3} \alpha^*_{i_3 j_3} \alpha^*_{i_4 j_4} \\ \nonumber
& = & \sum
\alpha^{i_1 j_1} \alpha^{i_2 j_2} \alpha^*_{i_1 j_2} \alpha^*_{i_2 j_1}
\end{array}
\end{equation}
For two qubits, we know that this is the only other independent invariant
because every state has a canonical Schmidt form $\Ket{\psi} = \sqrt{p} \Ket{00}
+ \sqrt{1-p} \Ket{11}$, with $1/2 \leq p \leq 1$ and $J = 2(p^2 - p) + 1$
determines $p$ uniquely.

\begin{figure}
\includegraphics[angle = 0, width = 8.3cm, height = 2.8cm]{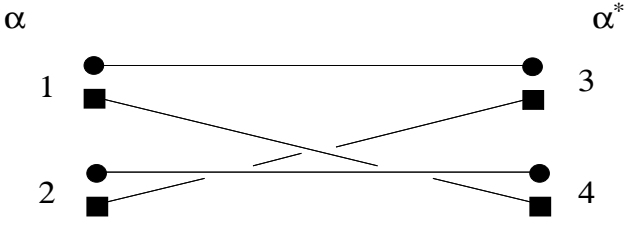}
\caption{\label{Poly:2qdiag}Diagrammatic representation of the quartic two-qubit
LU invariant J, given in eq.(\ref{Poly:2qPer}).  The first index of each term is
represented by a circle and the second by a square.  A line joins indices that are contracted with a $\delta$.}
\end{figure}

Another useful way of representing the invariant is to define two permutations
$\sigma,\tau$ on the set $\{1,2\}$ where $\sigma$ is the identity permutation
and $\tau(1) = 2, \tau(2) = 1$.  Then
\begin{equation}
\label{Poly:2qPer}
J_{(\sigma,\tau)} = \sum \alpha^{i_1 j_1} \alpha^{i_2 j_2} \alpha^*_{i_{\sigma(1)} j_{\tau(1)}} \alpha^*_{i_{\sigma(2)} j_{\tau(2)}}
\end{equation}
This also suggests a diagrammatic way of representing the invariant (see fig. \ref{Poly:2qdiag}).

\subsubsection{General case}

A multipartite pure state can be written in terms of an orthonormal
basis as follows
\begin{equation}
\Ket{\psi} = \sum_{i,j,k \ldots } \alpha^{ijk \ldots}
\Ket{ijk\ldots}
\end{equation}
A general polynomial function of the state coefficients and their complex
conjugates can be written as
\begin{equation}
\label{Poly:General}
\begin{array}{ll}
\sum & c_{i_1 j_1 k_1 \ldots i_2 j_2 k_2 \ldots}^{i_r j_r
k_r \ldots}
 \alpha^{i_1 j_1 k_1 \ldots} \alpha^{i_2 j_2 k_2 \ldots}
\ldots \\ \nonumber & \qquad \alpha^*_{i_r
j_r k_r \ldots}\ldots
\end{array}
\end{equation}
If the polynomial (\ref{Poly:General}) has equal numbers of $\alpha$'s and
$\alpha^*$'s and all the indices of the $\alpha$'s are contracted using the
invariant tensor $\delta$ with those of the $\alpha^*$'s, each index being
contracted with an index corresponding to the same party then the polynomial
is manifestly invariant under LU transformations.

Such polynomials can be written in terms of permutations on the
indices.  Let $r$ be the degree of the polynomial in $\alpha$ (and hence also
the degree in $\alpha^*$).  Let $\sigma,\tau, \mu \ldots$ be permutations acting
on the set $\{1,2,\ldots,r\}$ and let  $\vec{\sigma} = (\sigma, \tau, \mu,
\ldots)$. Then the invariants can be written as:
\begin{equation}
\label{Poly:PerDel}
\begin{array}{ll}
J_{\vec{\sigma}} = \sum & \alpha^{i_1 j_1 k_1 \ldots} \alpha^{i_2 j_2 k_2 \ldots} \ldots \\
\nonumber & \qquad \alpha^*_{i_{\sigma(1)} j_{\tau(1)} k_{\mu(1)}\ldots}
\alpha^*_{i_{\sigma(2)} j_{\tau(2)} k_{\mu(2)} \ldots} \ldots
\end{array}
\end{equation}
In fact, $\sigma$ can always be chosen to be the identity permutation by
permuting the $\alpha$ terms in this expression.  Additionally, each $J_{\vec{\sigma}}$ can be associated with a diagram constructed in the same way as fig.\ref{Poly:2qdiag}.

The invariants $J_{\vec{\sigma}}$ are enough to completely distinguish the
generic orbits under LU transformations.  In fact, invariant theory guarantees that only a
finite collection of them are needed to do this.  However, except in a few
simple cases, it is unknown which $J_{\vec{\sigma}}$ invariants form minimal
complete sets.

\subsection{Mixed states}

Two mixed states $\rho, \rho'$ are equivalent under LU transformations if
\begin{equation}
\rho' = U_1 \otimes U_2 \otimes \ldots \otimes U_n \rho U_1^\dagger \otimes U_2^\dagger \otimes \ldots \otimes U_n^\dagger
\end{equation}
The LU invariants for mixed states can be derived by rewriting the pure state
invariants (\ref{Poly:PerDel}) in terms of the density matrix $\rho =
\Ket{\psi}\Bra{\psi}$ and noting that the resulting expressions are still
invariant under LU transformations for general density matrices.  This can be
done by noting that terms such as $\alpha^{i_1 j_1 \ldots}\alpha^*_{i_2 j_2
\ldots}$ are elements of the density matrix.  A general density matrix may be
written in terms of an orthonormal basis as
\begin{equation}
\label{Poly:GenMixed}
\rho = \sum \rho^{ijk\ldots}_{mnp\ldots} \Ket{ijk\ldots} \Bra{mnp\ldots}
\end{equation}
and the corresponding expression for an LU invariant is
\begin{equation}
\label{Poly:PerMix}
\begin{array}{ll}
J_{\vec{\sigma}} = \sum & \rho_{i_{\sigma(1)} j_{\tau(1)}
k_{\mu(1)}\ldots}^{i_1 j_1 k_1 \ldots}  \rho_{i_{\sigma(2)}
j_{\tau(2)} k_{\mu(2)} \ldots}^{i_2 j_2 k_2 \ldots} \ldots \\
\nonumber & \qquad \rho_{i_{\sigma(r)} j_{\tau(r)} k_{\mu(r)} \ldots}^{i_r j_r
k_r \ldots}
\end{array}
\end{equation}

\section{Measuring Invariants under LU transformations}

\label{MeasLU}

\subsection{Network construction}

\label{MeasLU:Network}

\begin{figure}
\includegraphics[angle = 0, width = 8.3cm, height = 5.0cm]{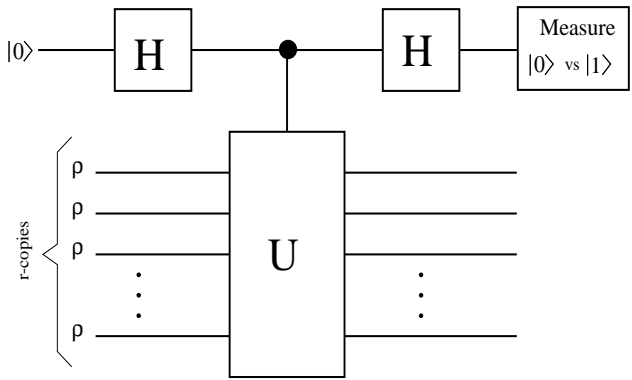}
\caption{\label{MeasLU:Net}General construction of network to measure polynomial
LU invariants.}
\end{figure}

The general construction of the network used to measure the LU invariants is
shown in fig.\ref{MeasLU:Net}.  It generalizes networks for estimating
functionals of bipartite states given in \cite{HorEkDirect, VarEst, HorEntEst}.  To measure an LU invariant of degree $r$ in
$\alpha$ (and also degree $r$ in $\alpha^*$) we take $r$ copies of the unknown
state $\rho$.   In addition, we take a single qubit in the state $\Ket{0}$ and
apply a Hadamard rotation $H$ to transform the state to $\frac{1}{\sqrt{2}}
\left ( \Ket{0} + \Ket{1} \right )$.  In the next step, we apply a unitary
operation $U$ on the $r$ copies of $\rho$ controlled by the Hadamard rotated
qubit.  Finally we perform a measurement on the single qubit in the $\{ \Ket{0},
\Ket{1}\}$ basis.  The expectation value of this measurement will be
\begin{equation}
\label{Meas:Expect}
\langle Z \rangle = \mbox{Re} \left ( \tr \left( U \rho^{\otimes r} \right) \right )
\end{equation}
When $\rho = \Ket{\psi}\Bra{\psi}$ is a pure state then this is equivalent to
\begin{equation}
\label{Meas:ExpectPure}
\langle Z \rangle = \mbox{Re} \Bra{\psi}^{\otimes r} U \Ket{\psi}^{\otimes r}
\end{equation}
In order to determine networks for measuring the LU invariants, it only remains
to show that there is a $U$ such that the invariants can be expressed in the
form (\ref{Meas:Expect}).

\begin{figure}
\includegraphics[angle = 0, width = 8.3cm, height = 3.8cm]{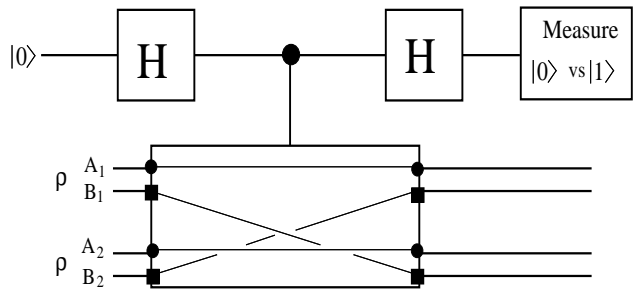}
\caption{\label{Meas:2qnet} Network for measuring the 2-qubit quartic invariant.}
\end{figure}

To do this for pure states, we have to express polynomials of the form (\ref{Poly:PerDel}) in
the form of (\ref{Meas:ExpectPure}).  Firstly, we note that $\alpha^{i j k
\ldots } = \BraKet{i j k \ldots }{\psi}$, $\alpha^*_{i j k \ldots} =
\BraKet{\psi}{i j k \ldots}$ and to each permutation $\sigma$ in
(\ref{Poly:PerDel}) we associate a permutation matrix
\begin{equation}
P_{\sigma} = \sum_{i_1,i_2,\ldots,i_r = 1}^d \Ket{i_{\sigma(1)} i_{\sigma(2)}
\ldots i_{\sigma(r)}} \Bra{i_1 i_2 \ldots i_r}
\end{equation}
where $P_{\sigma}$ acts on the Hilbert space of the same party for each of the $r$ copies of the state
$\Ket{\psi}$.  Then to each $\vec{\sigma}$ we associate the permutation matrix
\begin{equation}
P_{\vec{\sigma}} = P_{\sigma} \otimes P_{\tau} \otimes P_{\mu} \otimes \ldots
\end{equation}
where $P_{\sigma}, P_{\tau}, P_{\mu}, \ldots$ act on the Hilbert space of the same party as $\sigma, \tau,
\mu, \ldots$ in (\ref{Poly:PerDel}) on each of the $r$ copies of the state.  Then (\ref{Poly:PerDel}) can be written as
\begin{equation}
J_{\vec{\sigma}} = \Bra{\psi}^{\otimes r} P_{\vec{\sigma}} \Ket{\psi}^{\otimes r}
\end{equation}
Since $P_{\vec{\sigma}}$ is unitary these invariants can be estimated with the
network in fig.\ref{MeasLU:Net} by setting $U =
P_{\vec{\sigma}}$ to obtain the real part and $U = iP_{\vec{\sigma}}$ to obtain
the imaginary part.  For the specific example of the 2-qubit invariant
(\ref{Poly:2qqLU}) we have
\begin{equation}
\begin{array}{l}
J_{(\sigma,\tau)} = \Bra{\psi}_{A_1 B_1} \Bra{\psi}_{A_2 B_2} I_{A_1
A_2} \otimes SWAP_{B_1 B_2} \\ \qquad\qquad\qquad\qquad\qquad\qquad\qquad\Ket{\psi}_{A_1 B_1} \Ket{\psi}_{A_2 B_2}
\end{array}
\end{equation}
Note also that the physical construction of $P_{\vec{\sigma}}$ is closely
related to the diagram associated with $J_{\vec{\sigma}}$ (compare
figs. \ref{Poly:2qdiag} and \ref{Meas:2qnet} for example).

Finally, note that if $\rho$ is a mixed state then applying the same procedure
without modification will give the invariants of eq. (\ref{Poly:PerMix}).

It has previously been noted \cite{GrasslInv} that all homogeneous polynomial
LU invariants are determined by the expectation values of two observables on $r$
copies of a state.  Here, we have given an explicit network for measuring these
observables.  Also, similar constructions can be made to estimate other
polynomial functionals of quantum states \cite{VarEst} and these can be modified
to enable the estimation to proceed by LOCC \cite{VarLOCCEst}, i.e. with no
collective operations over the $n$-parties.  A similar modification would enable
the LU invariants to be estimated by LOCC, but this would affect the efficiency
of the estimation discussed in \S\ref{Eval:Comp}.

\section{Polynomial invariants under SLOCC}

\label{PolySLOCC}

When attempting to classify entanglement, it is often useful to consider
invariants under local transformations that are more general than unitary transformations.  For this purpose,
invariants under SLOCC have been introduced\cite{BenSLOCC}.  In \S\ref{MeasSLOCC} we construct a network to measure the
modulus squared of these invariants for the case where each party has a single qubit
(i.e. the Hilbert space is $(\mathbb{C}^2)^{\otimes n}$).

\subsection{Pure states}

Two $n$-party pure states $\Ket{\psi}$ and $\Ket{\psi'}$ are equivalent under SLOCC if it is
possible to obtain $\Ket{\psi'}$ with non-zero probability via a sequence of Local
Operations and Classical Communication (LOCC) starting from a single copy of
$\Ket{\psi}$ and vice-versa.  In \cite{Cirac3q}, this criterion was shown to be equivalent
to
\begin{equation}
\label{Poly:SLOCCPure}
\Ket{\psi'} =  M_1 \otimes M_2 \otimes \ldots \otimes M_n \Ket{\psi}
\end{equation}
where $M_j \in GL(d_j)$ is an invertible linear transformation acting on the
$d_j$-dimensional Hilbert space of the $j$th party.

In what follows, we find polynomial invariants for the special case where $M_j
\in SL(2)$, i.e. the transformation has unit determinant and each party has a
single qubit.  Networks to determine the modulus squared of these invariants
will be given in \S\ref{MeasSLOCC}. Note that it is not possible to measure
the $SL(2)^n$ invariants directly because they are not invariant under global
phase transformations $\Ket{\psi} \rightarrow e^{i \theta}\Ket{\psi}$, which
have no physical significance.  It is for this reason that we instead measure
the modulus squared, which is invariant under these phase transformations.

Under general $GL(2)^n$ transformations, the polynomial $SL(2)^n$ invariants are
still invariant up to a multiplicative factor, which is just some power of the
determinant of $M_1 \otimes M_2 \otimes \ldots \otimes M_n$.  Thus, ratios of
appropriate powers of these polynomials will be invariants under $GL(2)^n$.

\subsubsection{Two qubits}

In order to illustrate the polynomial invariants under $SL(2)^n$, first consider the case
where $n=2$.  Two states $\Ket{\psi} = \sum_{j,k = 1}^2 \alpha^{jk} \Ket{jk}$
and $\Ket{\psi'} = \sum_{j,k=1}^2 \alpha'^{jk} \Ket{jk}$ satisfy
(\ref{Poly:SLOCCPure}) if
\begin{equation}
\alpha' = M_1 \alpha M_2^T
\end{equation}
This means that $\det (\alpha) = \det (\alpha')$ is an $SL(2) \times SL(2)$
invariant, since $\det (M_1) = \det (M_2) = 1$.  This may be written as
\begin{equation}
\label{PolySLOCC:quad}
\det{\alpha} = \sum \epsilon_{i_1 i_2} \epsilon_{j_1 j_2} \alpha^{i_1 j_1} \alpha^{i_2 j_2}
\end{equation}
where the totally antisymmetric tensor $\epsilon_{ij}$ is the $SL(2)$ invariant
tensor.  For two qubit pure states, this is the only independent $SL(2) \times
SL(2)$ invariant.

\subsubsection{General case}

The $SL(2)^n$ invariants can be constructed in a similar way to the LU
invariants except the invariant tensor is now $\epsilon_{ij}$, and we contract
$\alpha$'s with $\alpha$'s instead of $\alpha^*$'s.  Thus, polynomials of the
form
\begin{equation}
\label{Poly:PerEp}
\begin{array}{ll}
K_{\vec{\sigma}} = \sum_1^2 & \epsilon_{i_1 i_2}\epsilon_{j_1 j_2} \epsilon_{k_1
k_2} \ldots  \epsilon_{i_{r-1} i_r}\epsilon_{j_{r-1} j_r} \epsilon_{k_{r-1} k_r}
\\ \nonumber & \qquad \alpha^{i_{\sigma(1)} j_{\tau(1)} k_{\mu(1)} \ldots}
\alpha^{i_{\sigma(2)} j_{\tau(2)} k_{\mu(2)} \ldots} \ldots \\ \nonumber &
\qquad \alpha^{i_{\sigma(r)} j_{\tau(r)} k_{\mu(r)}\ldots}
\end{array}
\end{equation}
are manifestly invariant.  Note that it is straightforward to generalize this
construction to the case where each party has a $d$-dimensional Hilbert space by
contracting with the $SL(d)^n$ invariant tensor $\epsilon_{i_1 i_2 \ldots i_d}$
instead of $\epsilon_{ij}$.  However, it is not yet clear how to measure these
invariants because the effect of the higher rank $\epsilon$ tensors cannot be
physically implemented by linear transformations on states.

\subsection{Mixed states}

In general, two mixed states $\rho, \rho'$ are equivalent under SLOCC if there
exists two completely positive maps $\mathcal{E}_1, \mathcal{E}_2$ which are
implementable via LOCC with non-zero probability of success such that $\rho' =
\mathcal{E}_1(\rho)$ and $\rho = \mathcal{E}_2(\rho')$.  In order derive invariants
using the expressions from the previous section, we will restrict to the case
where $\rho$ and $\rho'$ are related by
\begin{equation}
\rho' = M_1 \otimes M_2 \otimes \ldots \otimes M_n \rho M_1^\dagger \otimes
M_2^\dagger \otimes \ldots \otimes M_n^\dagger
\end{equation}
with $M_j \in SL(2)$.  The resulting expressions may not be invariant under
more general SLOCC transformations, but are related to important quantities in
entanglement theory as described in \S\ref{SLOCCExamples}

Unlike the LU invariants, it is not clear that (\ref{Poly:PerEp}) can be written
simply in terms of the coefficients of the density matrix $\rho = \Ket{\psi}\Bra{\psi}$.  However, $\left |
K_{\vec{\sigma}} \right |^2$ can be written as follows
\begin{equation}
\label{Poly:EpMix}
\begin{array}{ll}
\left | K_{\vec{\sigma}} \right |^2 = \sum_1^2 & \epsilon_{i_1 i_2}\epsilon_{j_1 j_2} \epsilon_{k_1 k_2} \ldots \epsilon_{i_{r-1} i_{r}} \epsilon_{j_{r-1}
j_r} \epsilon_{k_{r-1} k_r} \\
\nonumber & \epsilon^{m_1 m_2}\epsilon^{n_1 n_2} \epsilon^{p_1 p_2}
\ldots \epsilon^{m_{r-1} m_r} \epsilon^{n_{r-1}
n_r} \epsilon^{p_{r-1} p_r} \\
\nonumber &  \qquad \rho^{i_{\sigma(1)} j_{\tau(1)} k_{\mu(1)} \ldots}_{m_{\sigma(1)} n_{\tau(1)}
p_{\mu(1)} \ldots} \rho^{i_{\sigma(2)} j_{\tau(2)} k_{\mu(2)} \ldots}_{
m_{\sigma(2)} n_{\tau(2)} p_{\mu(2)} \ldots} \ldots \\ \nonumber & \qquad \rho^{i_{\sigma(r)}
j_{\tau(r)} k_{\mu(r)}\ldots}_{m_{\sigma(r)} n_{\tau(r)} p_{\mu(r)} \ldots}
\end{array}
\end{equation}
and these will also be $SL(2)^n$ invariants for mixed states.

\subsection{Examples of $SL(2)^n$ invariants}

\label{SLOCCExamples}

The $K_{\vec{\sigma}}$ invariants are especially interesting in entanglement theory
because many important entanglement measures can be easily calculated from them.
For example, in the case of two-qubits, the concurrence \cite{WootCon} is defined as a simple
function of the eigenvalues of $\rho \tilde{\rho}$, where
\begin{equation}
\label{Poly:2qtilde}
\tilde{\rho} = \sigma_y \otimes \sigma_y \rho^T \sigma_y \otimes \sigma_y
\end{equation}
and $^T$ stands for transpose in the computational basis.  These eigenvalues can be calculated from
$\tr((\rho \tilde{\rho})^m)$ for $m = 1,2,3,4$, which are simply the moduli
squared of $K_{\vec{\sigma}}$ invariants.  In \cite{HorEntEst}, networks were
constructed to estimate these invariants for two qubits and we will generalize
this construction to $K_{\vec{\sigma}}$ invariants for larger numbers of parties.

Another interesting example is the 3-tangle \cite{Woot3tang1,Woot3tang2}, which is defined for pure states as the modulus of the following 3-qubit $K_{\vec{\sigma}}$ invariant.
\begin{equation}
\begin{array}{ll}
\tau_{3} = \sum_1^2 & \alpha^{i_1 j_1 k_1} \alpha^{i_2 j_2 k_2} \epsilon_{i_1 i_3} \epsilon_{j_1 j_3}
\epsilon_{k_1 k_4} \epsilon_{i_2 i_4} \epsilon_{j_2 j_4} \epsilon_{k_2 k_3} \\
\nonumber & \alpha^{i_3 j_3 k_3} \alpha^{i_4 j_4 k_4}
\end{array}
\end{equation}
The 3-tangle gives information about the genuine 3-party entanglement between
the qubits.

\begin{figure}
\includegraphics[angle = 0, width = 8.3cm, height = 2.8cm]{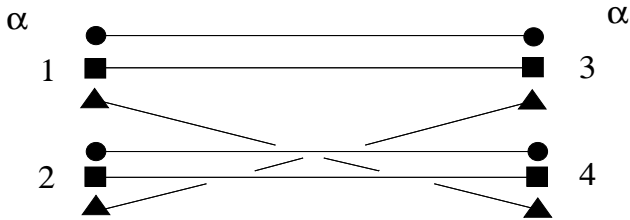}
\caption{\label{Poly:3tangdiag}Diagrammatic representation of the 3-tangle.  The
first index of each term is represented by a circle, the
second by a square and the third by a triangle.  A line joins indices that are contracted with an $\epsilon$.}
\end{figure}

Finally, note that the $K_{\vec{\sigma}}$ invariants can be given similar diagrammatic
representations to the $J_{\vec{\sigma}}$ invariants.  This is illustrated for the
3-tangle in fig.\ref{Poly:3tangdiag}.

\section{Measuring SLOCC invariants}

\label{MeasSLOCC}

The modulus squared of the SLOCC invariants can be measured using a network
similar to fig. \ref{MeasLU:Net} except that the unknown states $\rho$ must be
preprocessed prior to the controlled-$U$ operation.  If $K_{\vec{\sigma}}$ is of
degree $r$ in $\alpha$ then we will need $r$ copies of $\rho$.  The
preprocessing stage will consist of collective unitary operations and completely
positive maps that act on the entire Hilbert space of the $r$ copies of $\rho$.
The resulting state $\rho'$, will yield the expectation value
\begin{equation}
\langle Z \rangle = \mbox{Re} \left ( \tr \left ( U \rho' \right ) \right )
\end{equation}
for the measurement at the end of the network.  In this section, we describe the
preprocessing operations and unitary operations $U$ that enable the modulus
squared of the SLOCC invariants to be written in this form.

First, we apply the inverse of the permutation matrix associated with
$\vec{\sigma}$ to the $r$ copies of $\rho$ to obtain $P_{\vec{\sigma}}^\dagger
\rho^{\otimes r}P_{\vec{\sigma}}$.

The second, and final, part of the preprocessing stage is to apply a completely
positive map $\bar{\Lambda}$ to the state. To describe $\Lambda$ we first define
the multi-party analogue of eq. (\ref{Poly:2qtilde}).
\begin{equation}
\tilde{\rho} = \sigma_y \otimes \sigma_y \otimes \ldots \otimes \sigma_y \rho^T \sigma_y \otimes \sigma_y \otimes \ldots \otimes \sigma_y
\end{equation}
Next, we define a map $\Lambda$ that acts on a product of $r$ states
by applying the tilde operation to the even numbered states as follows
\begin{equation}
\Lambda(\rho_1 \otimes \rho_2 \otimes \ldots \otimes \rho_r) = \rho_1 \otimes \tilde{\rho}_2 \otimes \rho_3 \otimes \ldots \otimes \tilde{\rho}_r
\end{equation}
where each $\rho_j$ is an $n$-party state.

Unfortunately, $\Lambda$ cannot be physically implemented, since it is not a
completely positive map.  This can be dealt with by using the Structural
Physical Approximation (SPA) to $\Lambda$, which we will call $\bar{\Lambda}$.
$\bar{\Lambda}$ is the ``closest'' physical map to $\Lambda$.  This is discussed in \S\ref{SPA}, but for now we construct the network as if $\Lambda$ could be implemented perfectly.

The final pre-processed state $\rho'$ will be
\begin{equation}
\rho' = \Lambda(P^{\dagger}_{\vec{\sigma}}\rho^{\otimes r} P_{\vec{\sigma}})
\end{equation}

Next, the controlled-$U$ operation in our network must be chosen such that
$\langle Z \rangle = |K|_{\vec{\sigma}}^2$ when $\rho'$ is used as the input.
One can easily verify that the pairwise SWAP gate, defined by
\begin{equation}
\label{Meas:Shift}
\begin{array}{c}
U \Ket{\phi_1}\otimes\Ket{\phi_2}\otimes \ldots \otimes \Ket{\phi_{r-1}}
\otimes \Ket{\phi_{r}} = \\ \Ket{\phi_{2}} \otimes \Ket{\phi_1} \otimes
 \ldots \otimes \Ket{\phi_r} \otimes \Ket{\phi_{r-1}}
\end{array}
\end{equation}
where $\Ket{\phi_j}$ is an $n$-party state fulfils this condition.

\section{The Structural Physical Approximation}

\label{SPA}

The $\Lambda$ operation encountered in the previous section is an example of a
positive, but not completely positive map.  These cannot be implemented exactly,
but instead we can apply an approximation.
\begin{equation}
\bar{\Lambda} \left ( \rho \right ) = \alpha I + \beta \Lambda(\rho)
\end{equation}
where $I$ is the identity operator and $\alpha, \beta$ are real positive
constants chosen such that $\bar{\Lambda}$ is completely positive.  If we fix
$\alpha$ and $\beta$ such that $\bar{\Lambda}$ is trace-preserving and $\beta$
is maximized, then the results of \cite{HorSPA} imply that
\begin{equation}
\bar{\Lambda} \left ( \rho \right ) = \frac{2^{\frac{3}{2}nr}}{2^{\frac{3}{2}nr} + 1}
\frac{I}{2^{nr}} + \frac{1}{2^{\frac{3}{2}nr} + 1} \Lambda \left ( \rho \right )
\end{equation}
where $n$ is the number of qubits in each copy of the state and $r$ is the
degree of the $K_{\vec{\sigma}}$ for which we are estimating the modulus
squared.

On replacing $\Lambda$ with $\bar{\Lambda}$ in
our network the expectation value of the $Z$ measurement still allows the modulus squared of
the $K_{\vec{\sigma}}$ invariant to be determined via
\begin{equation}
\label{Meas:SPAEst}
\left | K_{\vec{\sigma}} \right |^2 = \left ( 2^{\frac{3}{2}nr} + 1 \right ) \left < Z \right > - 2^{nr}
\end{equation}
However, the SPA does affect the accuracy to which the invariant is
determined. This is discussed further in the next section.  Additionally, in
\cite{VarLOCCEst}, it is shown that this sort of SPA can be implemented by
LOCC.  Thus, the SLOCC invariants could also be estimated by LOCC, but the
efficiency discussed in \S\ref{Eval:Comp} would be affected.

\section{Evaluation}

\label{Eval}

The main aim of the protocols presented in \S\ref{MeasLU} and \S\ref{MeasSLOCC} is to provide a
physical interpretation for the polynomial invariants.  However, we have not yet
addressed the question of how efficient these measurement protocols are.  In
this section, we compare the efficiency of our protocols to protocols based on simply measuring the
state coefficients and calculating the invariants.  We use unbiased estimators
based on counting \cite{SmithTom, LeonTom, dArTom}.  Also, we perform the
analysis in the limit where a large number of copies of the state have been
measured, so that the variances of the estimates are small and can be treated to
first order in all subsequent calculations.  We note that more sophisticated
estimation procedures are also possible \cite{GillTom}, but our purpose here is
to compare the networks to methods that are easily accessible experimentally.

Measuring the state coefficients would clearly be a more straightforward
procedure to perform experimentally than using our network.  Although more
parameters have to be determined, this does not necessarily mean that it is a
less efficient method for estimating the invariants than using our networks.
There are several quite general reasons why this might be the case.

Firstly, suppose that we are interested in measuring a complete set of
polynomial LU invariants for some unknown state of $n$ parties, where each
party has a $d$-dimensional Hilbert space.  In general, we do not know how
many we would need to measure, but parameter counting arguments \cite{Multi1,
Multi2, Multi3} show that the number of local degrees of freedom is linear in
$n$ whereas the total number of degrees of freedom is exponential in $n$.  Thus,
for large $n$ almost all the degrees of freedom are non-local.  Even for moderately sized $n$, there are nearly as many
invariants as there are state coefficients.  In addition, the invariants are typically
highly non-linear functions of the state coefficients.  For these reasons, we
expect that measuring a complete set of invariants directly will generally not be more
efficient than measuring the state coefficients for large $n$.  Similar considerations also
apply to the SLOCC invariants.

Despite these considerations, it may be the case that our networks are more
efficient if we are only interested in measuring a small incomplete subset of
the invariants.  Also, they may be more efficient for estimating complete sets
when $n$ is small.  For this reason, and for simplicity, we concentrate on
estimating two qubit invariants in this section.

There are also other reasons why our protocols may not be efficient.  For
example, our protocols only employ a two-outcome measurement for each $r$ copies of
the state whereas estimating the state coefficients uses a two-outcome
measurement on each copy.  Also, for the $K_{\vec{\sigma}}$ invariants, we will
see that using the SPA introduces a lot of noise into the measurement.
Nonetheless, there are still some cases where using our networks is more
efficient than estimating the state coefficients.

\subsection{Statistical analysis of the network}

\label{Eval:Stats}

For a particular setup in our network we make repeated measurements of an
observable $Z$, with expectation value $F = \tr \left ( U \rho' \right )$.  $Z$
is a random variable\footnote{The statistical inference theory used in this
section can be found in many statistics textbooks, such as \cite{CoxHinkley}.} with distribution
\begin{equation}
\begin{array}{l}
p(Z = +1) = \frac{1}{2} \left ( 1 + F \right ) \\
p(Z = -1) = \frac{1}{2} \left ( 1 - F \right )
\end{array}
\end{equation}
If we define the event $Z = +1$ as a success and set $p = P(Z=+1)$ then repeating
the network $N$ times is equivalent to performing $N$ Bernoulli trials.  The
number of successes $N_s$ is a random variable with a binomial distribution
and its expectation value is $\left < N_s \right > = Np = \frac{N}{2} \left
( 1 + F\right )$.  In an actual experiment, the observed number of successes $\hat{N_s}$ can be used to
compute an unbiased estimator for $F$, given by
\begin{equation}
\label{Eval:FEst}
\hat{F} = 2 \frac{\hat{N_s}}{N} - 1
\end{equation}
with variance
\begin{equation}
\var \left ( \hat{F} \right ) = \frac{1}{N} \left ( 1 - F^2\right )
\end{equation}
We are interested in determining how many trials are needed in order for the
estimate $\hat{F}$ to be reasonably accurate.  Specifically, we would like to
quantify how many trials are needed to make the variance of $\var ( \hat{F} ) \leq
\epsilon$ for some $\epsilon > 0$.  In an experimental situation, we would not
be able to calculate $\var ( \hat{F} )$ from our data, so we would have to estimate
it using the sample variance, $\hat{\var} ( \hat{F})$.  However, in the limit $N
\rightarrow \infty$ we can use the fact that $\var ( \hat{F} ) = O(N^{-1})$ and
$\var ( \hat{\var} (\hat{F})) = O(N^{-4})$, i.e. $\hat{\var} ( \hat{F})$ converges to
the true variance much faster than $\hat{F}$ converges to $F$ so $\hat{\var} ( \hat{F}) \approx \var ( \hat{F} )$.  Thus, in this limit we have that
\begin{equation}
N \gtrapprox \frac{1}{\epsilon} (1 - F^2)
\end{equation}

Recall that for the LU invariants, the real and imaginary parts of the invariant
are estimated independently and that each use of the network requires $r$ copies
of the state, where $r$ is the degree of the invariant in $\alpha$.  If we use the same number of samples for estimating both the real and imaginary parts then the total number of copies required is
\begin{equation}
M \gtrapprox \frac{r}{\epsilon} \left ( 2 - |J_{\vec{\sigma}}|^2 \right )
\end{equation}
In some cases, we know a priori that the invariant is always real or always
imaginary. If this is the case, then we can achieve the same accuracy with
\begin{equation}
\label{Eval:NetComp}
M \gtrapprox \frac{r}{\epsilon} \left ( 1 - |J_{\vec{\sigma}}|^2 \right )
\end{equation}

For the SLOCC invariants, each use of the network requires $r$ copies of the
state, where $r$ is the degree of the invariant in $\alpha$.  Also the estimate
of the invariant must take into account the use of the SPA via
(\ref{Meas:SPAEst}).  In this case, the total number of copies required is
\begin{equation}
\label{Eval:SLOCCNet}
M \gtrapprox \frac{r}{\epsilon} \left [ \left ( 2^{\frac{3}{2}nr} + 1 \right )^2
- \left ( |K_{\vec{\sigma}}|^2 + 2^{nr} \right )^2 \right ]
\end{equation}
Notice that the $2^{3nr}$ term will dominate the term in the square bracket for
large $n$ and $r$.  This is due to the noise introduced into the measurement by
the SPA.

\subsection{Comparison to methods based on state estimation}

\label{Eval:Comp}

In order to evaluate our protocols, we compare them to methods based on
estimating the density matrix of the state and then calculating the invariants.
We do this by estimating each state coefficient using observations on single
copies of the state.  This is known as homodyne tomography (see \cite{GillTom}
for an overview and also \cite{SmithTom,LeonTom,dArTom}). This is not the
optimal way of reconstructing the state in general\cite{GillStat}, but it will
greatly simplify the analysis.

\subsubsection{Example: Two-qubit LU invariants}

A general two-qubit density matrix can be written as
\begin{equation}
\label{Eval:2qrho}
\begin{array}{ll}
\rho = \frac{1}{4} & \left ( I_2 \otimes I_2 + \sum_j a_j \sigma_j \otimes
I_2 + \sum_j b_j I_2 \otimes \sigma_j \right. \\ \nonumber & \left. + \sum_{j,k} R_{jk} \sigma_j \otimes
\sigma_k \right )
\end{array}
\end{equation}
The two-qubit LU invariant (\ref{Poly:2qqLU}) can be written in terms of these
coefficients as
\begin{equation}
J = \tr (\rho_B^2) = \frac{1}{2} \left ( 1 + \sum_j b_j^2 \right )
\end{equation}

Each $b_j$ can be determined by simply performing a $\sigma_j$ measurement
on $N_j$ copies of Bob's half of the state.  The probability distributions of
the associated random variables are given by
\begin{equation}
\begin{array}{l}
p(\sigma_j = + 1) = \frac{1}{2} \left ( 1 + b_j \right ) \\
p(\sigma_j = - 1) = \frac{1}{2} \left ( 1 - b_j \right )
\end{array}
\end{equation}

Thus, each $b_j$ can be estimated in the same way as $F$ in (\ref{Eval:FEst})
and we have that
\begin{equation}
\var ( \hat{b_j} ) = \frac{1 - b_j^2}{N_j}
\end{equation}
We can then construct an estimator for $J$ given by
\begin{equation}
\hat{J} = \frac{1}{2} \left ( 1 + \sum_j \hat{b_j}^2 \right )
\end{equation}
which will be biased, but in the large $N_j$ limit
\begin{equation}
\var ( \hat{J} ) \approx  \sum_{j} b_j^2 \left ( \frac{1 - b_j^2}{N_j}\right )
\end{equation}
to first order in $\var (b_j)$.

If we make the additional restriction that each observable $\sigma_{j}$ is
sampled the same number of times (i.e. $N_j = \frac{N}{3}$) then we must take
\begin{equation}
\label{Eval:LUComp}
N \gtrapprox \frac{3}{\epsilon} \sum_j b_j^2 \left (1- b_j^2 \right )
\end{equation}
for our estimate to have variance $\lessapprox \epsilon$.

One way to compare this to the result for our network is to take an average over
all pure states.  If we assume that all pure states are equally likely,
i.e. integrate (\ref{Eval:Comp}) and (\ref{Eval:LUComp}) using Haar measure (for details see \cite{BuzekRecon}), then we find that on average we will need
$3/2$ times as many copies of the state if we use the coefficient estimation
method.  This is half of what one might expect from parameter counting alone,
since three times as many parameters are estimated in the state coefficient
method.  The factor of two is explained by the fact that each use of our network
uses two copies of the state.

However, it is possible to find parameter ranges in which the state coefficient
method performs better than our networks.  One such range is given by setting
$b_1 = b_2 = 0, -\sqrt{\frac{3}{5}} < b_3 < \sqrt{\frac{3}{5}}$.  This
illustrates the fact that parameter counting does not always reflect the
statistical efficiency of a given protocol.  Any partial information we have
available about the type of states being measured might change our judgement of
which protocol is more efficient.

\subsubsection{Example: Two-qubit SLOCC invariants}

For the two-qubit SLOCC invariants we take the quadratic invariant
(\ref{PolySLOCC:quad}) as an example.  In terms of the decomposition
(\ref{Eval:2qrho}) this can be written as
\begin{equation}
|K|^2 = \frac{1}{4} \left [ 1 - \sum_j \left ( a_j^2 + b_j^2 \right ) + \sum_{jk} R_{jk}^2 \right ]
\end{equation}
If we estimate this by measuring all $15$ of the state coefficients an equal number of
times then by a similar analysis to the LU case we find that we need at least
\begin{equation}
\label{Eval:SLOCCCoeff}
\begin{array}{ll}
N \gtrapprox \frac{15}{4\epsilon} & \left [ \sum_j \left [ a_j^2 (1 - a_j^2) +
b_j (1 - b_j^2 )\right ] \right. \\ \nonumber & \left. + \sum_{jk}
R_{jk}^2 (1 - R_{jk}^2) \right ]
\end{array}
\end{equation}
copies of the state to get a variance $\lessapprox \epsilon$.

Taking averages, one finds that fewer copies are needed in the state coefficient
protocol by a factor $\approx 5 \times 10^{3}$ despite the fact that many more
parameters have to be estimated in this protocol than when using our network.
This is largely due to the factor $2^{12}$ that appears in (\ref{Eval:SLOCCNet}),
which arises from the noise introduced by the SPA.  This suggests that other
estimation and detection protocols based on the SPA \cite{HorEkDirect,
HorEntEst} may be less efficient than parameter counting arguments would imply.
In fact, there are no states for which our network performs better than the
coefficient estimation method.  Even in the best possible case for our network,
the state coefficient method requires fewer states by about $3$ orders of
magnitude. 

\section{Conclusions}

\label{Conc}

We have presented networks for measuring the polynomial invariants of quantum
states under LU and SLOCC transformations.  The structure of these networks is
closely related to the structure of the invariants themselves and thus gives the
invariants a physical interpretation.  Comparison of these networks with methods
based on estimating the state coefficients indicate that the networks are of
limited practical use for estimating complete sets of invariants.  Indeed, our
results suggest that any estimation procedure that employs the SPA is
statistically inefficient even when the number of parties is small.

We know that no procedure for estimating invariants directly can outperform
protocols based on estimating the state coefficients as the number of parties
is increased.  For small numbers of parties it seems that there can be some
increase in efficiency, but the optimal protocol is not known in general.

\end{document}